\documentclass{aa}
\usepackage{graphicx}
\usepackage[varg]{txfonts}
\usepackage{natbib}
\usepackage{gensymb}
\usepackage{multirow}
\usepackage{color}
\usepackage[normalem]{ulem}

\bibpunct{(}{)}{;}{a}{}{,} 
\begin{document}

\title{Chemodynamics of metal-poor wide binaries in the Galactic halo: Association with the Sequoia event}

\author{Dongwook Lim\inst{1}
  \and Andreas J. Koch-Hansen\inst{1}
  \and Camilla Juul Hansen\inst{2}
  \and Sebastien L\'epine\inst{3} 
  \and Jennifer L. Marshall\inst{4}
  \and Mark I. Wilkinson\inst{5}
  \and Jorge Pe{\~n}arrubia\inst{6, 7}
  }

\authorrunning{D. Lim et al.}
\titlerunning{Metal-poor wide binaries}


\institute{Zentrum f\"ur Astronomie der Universit\"at Heidelberg, Astronomisches Rechen-Institut, M\"onchhofstr. 12-14, 69120 Heidelberg, Germany, 
  \email{dongwook.lim@uni-heidelberg.de}
  \and Max Planck Institute for Astronomy, K\"onigstuhl 17, 69117 Heidelberg, Germany
  \and Department of Physics and Astronomy, Georgia State University, 25 Park Place, Suite 605, Atlanta, GA 30303, USA
  \and George P. and Cynthia Woods Mitchell Institute for Fundamental Physics and Astronomy, and Department of Physics and Astronomy, Texas A\&M University, College Station, TX 77843, USA
  \and Department of Physics and Astronomy, University of Leicester, University Road, Leicester LE1 7RH, UK
  \and Institute for Astronomy, University of Edinburgh, Royal Observatory, Blackford Hill, Edinburgh EH9 3HJ, UK
  \and Centre for Statistics, University of Edinburgh, School of Mathematics, Edinburgh EH9 3FD, UK
  }

\date{Received July 7, 2021 / Accepted August 11, 2021}

\abstract {
Recently, an increasing number of wide binaries has been discovered. Their chemical and dynamical properties are  studied through extensive surveys and pointed observations.  
However, the formation of these wide binaries is far from clear, although several scenarios have been suggested. 
In order to investigate the chemical compositions of these systems,  we analysed high-resolution spectroscopy of three wide binary pairs belonging to the Galactic halo. 
In total, another three candidates from our original sample of 11 candidates observed at various resolutions with various instruments were refuted as co-moving pairs because their radial velocities are significantly different.
Within our sample of wide binaries, we found homogeneity amongst the pair components in dynamical properties (proper motion and line-of-sight velocities) and also in chemical composition.  
Their metallicities are $-$1.16, $-$1.42, and $-$0.79 dex in [Fe/H] for each wide binary pair, which places these stars on the metal-poor side of  wide binaries reported in the literature. 
In particular, the most metal-poor pair in our sample (WB2 $\equiv$ HD134439/HD134440) shows a lower [$\alpha$/Fe] abundance ratio than Milky Way field stars, which is a clear signature of an accreted object. 
We also confirmed that this wide binary shares remarkably similar orbital properties with stars and globular clusters associated with the Sequoia event. 
Thus, it appears that the WB2 pair was formed in a dwarf galaxy environment and subsequently dissolved into the Milky Way halo. 
Although the other two wide binaries appear to arise from a different formation mechanism, our results provide a novel opportunity for understanding the formation of wide binaries and the assembly process of the Milky Way. 
} 

\maketitle


\section{Introduction} \label{sec:intro}

\begin{table*}
\tiny 
\setlength{\tabcolsep}{4pt}
\caption{Target information}
\label{tab:target} 
\centering                                    
\begin{tabular}{c c c c c c c c c c c c}  
\hline\hline        
\multirow{2}{*}{ID}     & \multirow{2}{*}{Name} & \multirow{2}{*}{Gaia source ID}     & $\alpha$ (J2000)      & $\delta$      (J2000) & Parallax      & $\mu_{\alpha}$  & $\mu_{\delta}$        & 
        $G$             & RV                            & Ang. Sep      & Phy. Sep \\
                                &                                       &                                                         & [deg]                 & [deg]                   & [mas] & [mas yr$^{-1}$]       & [mas yr$^{-1}$]       & 
        [mag]   & [km~s$^{-1}$]         & [$\arcsec$]   & [AU] \\
\hline                                   
WB1a & --                       & 4012137072424462464 & 186.992602      & 31.153772       & 4.0615                & 14.982                & $-$175.106    & 12.52   & $-$70.88      &  \multirow{2}{*}{45.43} & \multirow{2}{*}{11,178}         \\
WB1b & --                       & 4012137106784200064 & 186.977891      & 31.154626       & 4.0660                & 15.026                & $-$174.778    & 12.63   & --                    & & \\
WB2a & HD134439 & 6307374845312759552 & 227.554530      & -16.379410    & 34.0175         & $-$998.064    & $-$3542.267   &  8.83 & 310.88                & \multirow{2}{*}{300.65} & \multirow{2}{*}{8,837}        \\
WB2b & HD134440 & 6307365499463905536 & 227.554033      & -16.462923    & 34.0266         & $-$1000.844   & $-$3540.080   &  9.16 & 311.35                & & \\
WB3a & --               & 2135974995372447232 & 293.664197      & 50.658265     & 4.8253          & 43.079                & 172.748               & 11.29 & $-$66.56        & \multirow{2}{*}{28.88}        & \multirow{2}{*}{5,977}        \\
WB3b & --               & 2135977950309943424 & 293.654464      & 50.653139     & 4.8376          & 42.549                & 172.984               & 11.24 & $-$64.67        & & \\
\hline                                             
\end{tabular}
\end{table*}

The existence of wide binaries that show wide separations between the two component stars from a few AU to more than several thousands AU is one of the important and unresolved topics in Galactic astronomy. 
Not only is the formation mechanism of these wide binaries puzzling, these objects are also an efficient tool for studying the halo structure and dark matter characteristics, as well as for testing the validity of chemical tagging  \citep[][and references therein]{Bahcall1985,Yoo2004,Chaname2004, Quinn2009,Monroy-Rodriguez2014, Hawkins2020, Tian2020, Hwang2021}.
By rights, common-motion pairs at separations of 1 pc should not exist because their wide separations make them susceptible to fast dissolution during encounters with even small substructures in the Milky Way halo.

Recent high-precision proper motion and parallax data provided by the first and second $Gaia$ data releases \citep{GaiaCollaboration2016, GaiaCollaboration2018} have enabled the discovery of an increasing number of co-moving pairs in the Milky Way \citep[e.g.][]{Andrews2017, Jimenez-Esteban2019, Hartman2020}. 
Because these co-moving pairs are only based on the proper motion, further confirmation of the common line-of-sight velocity is necessary to determine whether they are physically associated wide binary pairs, however.
In this regard, various spectroscopic surveys and in-depth observations are confirming the presence of wide binaries and reveal their chemical properties. 
For instance, \citet{Andrews2018} reported an identical metallicity of  wide binary components from the Radial Velocity Experiment (RAVE; \citealt{Kunder2017}) and the Large Sky Area Multi-Object Fibre Spectroscopic Telescope (LAMOST; \citealt{Luo2015}) data. \citet{Hawkins2020} found 
chemical homogeneity amongst the components of 20 pairs of wide binaries for a large number of elements, including $\alpha$ and neutron-capture elements.

Several scenarios have been suggested for the formation of wide binaries, such as formation during the early dissolution phase of young star clusters \citep{Kouwenhoven2010, Moeckel2011}, turbulent fragmentation \citep{Lee2017},
the dynamical unfolding of higher-order systems \citep{Reipurth2012, Elliott2016}, and formation from adjacent pre-stellar cores \citep{Tokovinin2017}.
More recently, \citet{Penarrubia2021} suggested that wide binaries could be formed in tidal streams of stars and globular clusters over a long period of time. 
Nevertheless, the formation of wide binaries is yet to be fully understood because most scenarios predict that the components of wide binaries should have similar chemical compositions, as is indeed observed.

On the other hand, detailed chemical abundance studies for various elements can reveal peculiar signatures of stars that trace back to their formation or evolution environment.
For instance, a low [$\alpha$/Fe] abundance ratio in moderately metal-poor stars implies that they were formed in a low-mass environment with a low star-forming efficiency, such as a dwarf galaxy \citep{Nissen2010}, while enhanced Na and Al, accompanied by O and Mg depletion in stars in the Milky Way indicates an origin in a globular cluster environment \citep{Fernandez-Trincado2017, Koch2019a, Lim2021}. 
Depleted [Mn/Fe] abundance ratios of stars in the Sagittarius dwarf galaxy and $r$-process enhancement of stars in the $Gaia$ Sausage/Enceladus are also reported as typical chemical signatures of accretion events \citep[see e.g.][]{McWilliam2003, Venn2004, Aguado2021, Koch-Hansen2021, Nissen2021}. 
In this regard, high-resolution spectroscopy for wide binaries can provide an opportunity to examine their birth environment and formation mechanism through detailed chemical abundance patterns when compared with stars in the various substructures of the Milky Way.  

We investigate the chemical composition of three wide binaries that belong to the Galactic halo using high-resolution spectroscopic data obtained from the 2.56 m Nordic Optical Telescope (NOT) on La Palma, Spain. This paper is organised as follows. In Sections~\ref{sec:target_obs} and \ref{sec:spec}, we describe the target selection, observation, data reduction process, and spectroscopic analysis. Our results for the chemical composition of wide binaries are presented in Section~\ref{sec:result}. Finally, we discuss the origin of these wide binaries in Section~\ref{sec:discussion}. 
In an appendix, we provide a list of common proper motion pairs that were refuted as physical binaries based on our derived differing radial velocities.


\section{Observations and data reduction} \label{sec:target_obs}
\subsection{Target selection and binary separations} \label{sec:target}
We have selected five wide binary candidates among stars with high proper motions ($\mu$ $>$ 40 mas~yr$^{-1}$) identified in the SUPERBLINK proper motion survey \citep{Lepine2005, Lepine2011}.
Halo candidates were first selected based on their location in a reduced proper motion diagram, indicative of a high transverse motion. 
We then searched for pairs of stars with angular separations $ \rho < 15' $ and proper motion differences $ \Delta \mu <$ 10 mas yr$^{-1}$, ultimately selecting 153 common proper motion pairs that based on pre-Gaia photometric distance estimates and assuming the pairs to be physical binaries would have projected orbital separations between 5,000 AU and 100,000 AU.
Our targets for follow-up spectroscopy were chosen for a compromise magnitude range between the bright primary and the usually much fainter secondary candidate, as well as for observability. 

The basic information for our target stars, updated by $Gaia$ eDR3 \citep{GaiaCollaboration2021}, is given in Table~\ref{tab:target}. 
In particular, the components of our WB2, that is, HD134439 and HD134440, are two of the fastest-moving objects on the sky as a well-known wide binary pair \citep{King1997, Allen2007}.
The angular separation for each wide binary pair is derived from equatorial coordinates, while the projected physical separation is converted using the angular separation and mean parallax of the two component stars.
The physical separations of our sample wide binaries are on the high side of known wide binaries \citep[$s$ $>$ 5,000 AU; see][]{Andrews2017, El-Badry2018, Tian2020}, although some ultra-wide binaries with separations $s$ $\gtrsim$ 0.1 pc have also been reported \citep{Hartman2020}. 

We further estimated the 3D separation adapting $Gaia$ eDR3 parallaxes, which are 57310$\pm$181810 AU, 8980$\pm$1190 AU, and 108850$\pm$108800 AU for the WB1, WB2, and WB3 pairs, respectively.
The 3D separation is much larger than the projected separation for the WB1 and WB3 pairs, while this difference is small for WB2.
However, these values might be partially due to the uncertainties of parallax measurement depending on the distance of the stars. 
More precise distance information to determine the 3D separation will be beneficial to investigate the Galactic gravitational potential because the wide binary fraction depending on its separation is one of the crucial tracers for mass constraints of the so-called massive compact halo objects (MACHOs, \citealt{Chaname2004, Quinn2009,Tian2020}).

We note that our initial list of follow-up targets consisted of five pairs, but two of them turned out to have very different radial velocities and were thus found to be chance alignments of unrelated objects.
These are not considered in the further analysis, but we list their details in Table~\ref{tab:append} in Appendix~\ref{apen:other}.

\subsection{FIES observations and data reduction} \label{sec:obs}
High-resolution spectra of the three wide binary pairs were obtained using the FIbre-fed Echelle Spectrograph (FIES, \citealt{Telting2014}) at the NOT on April 6 and 12, 2016, and on July 5, 2016.
We used `F3 (med-res) mode' with bundle C, which provides spectral coverage of 3640 -- 7260 $\AA$ at a spectral resolution $R \sim 45,000$. 
The total exposure time for each star is 1920, 2928, 80, 120, 608, and 652 seconds for WB1a, WB1b, WB2a, WB2b, WB3a, and WB3b, which were typically split into three exposures to facilitate cosmic-ray removal. 
Our data were reduced with the FIESTool pipeline\footnote{\url{http://www.not.iac.es/instruments/fies/fiestool/FIEStool.html}}, which performs standard reductions steps such as flat fielding and wavelength calibrations using Th-Ar lamp spectra.
Overall, our observation and reduction strategy yielded signal-to-noise ratios (S/N) of 20 -- 30 per pixel at around 5500 $\AA$.

Radial velocities of each star were determined by cross-correlation of the observed spectra against a template spectrum obtained from the POLLUX database\footnote{\url{http://pollux.oreme.org}} \citep{Palacios2010} using the {\em fxcor} task within IRAF RV package. 
The derived heliocentric velocities and measurement errors of stars are given in Table~\ref{tab:param}, which are similar to those provided by $Gaia$ (see Table~\ref{tab:target}).
The associated stars in each wide binary pair agree well in the line-of-sight velocity ($<$ 1.0 km s$^{-1}$), compared to the measurement error.
Together with their common proper motion, this close similarity strengthens the hypothesis that each of our targets is a physically bound co-moving wide binary system.   


\section{Spectroscopic analysis} \label{sec:spec}
\subsection{Atmospheric parameters}
We determined the stellar atmospheric parameters using a standard photometric approach. 
First, we calculated the effective temperature ($T_{\rm eff}$) from the (BP$-$RP) colour of $Gaia$ DR2 using the relation and coefficients of \citet{Mucciarelli2020}. 
The initial guess of [Fe/H] for this calculation was obtained from the code ATHOS \citep{Hanke2018}.
The reddening was derived from the 3D extinction map of Bayestar17 \citep{Green2018} with extinction coefficients from \citet{Casagrande2018}.
We estimated the bolometric correction using the code provided by \citet{Casagrande2018}.

Next, we calculated surface gravities ($\log{g}$) based on the canonical relation 
\small
\begin{eqnarray}
{\log{g_{*}} = \log{g_{\odot}} + \log{\frac{M_{*}}{M_{\odot}}} + 4\log{\frac{T_{\rm eff,*}}{T_{\rm eff,\odot}}} + 0.4(M_{bol,*} - M_{bol,\odot}),} 
\end{eqnarray}
\normalsize
where $\log{g_{\odot}}$ = 4.44 dex, $T_{\rm eff,\odot}$ = 5777 K, and $M_{bol,\odot}$ = 4.74 for the Sun. 
We assumed stellar masses of 1.15 $M_{\odot}$, 1.00 $M_{\odot}$ for WB1ab, 0.80 $M_{\odot}$ for WB2ab, and 1.20 $M_{\odot}$ for WB3ab based on the spectral types listed in the SIMBAD by comparing with the spectral standard of FGK main-sequence stars. 
The uncertainties in the $\log{g}$ determination from mass assumption are estimated in Section~\ref{sec:error}.

\begin{figure}
\centering
   \includegraphics[width=0.43\textwidth]{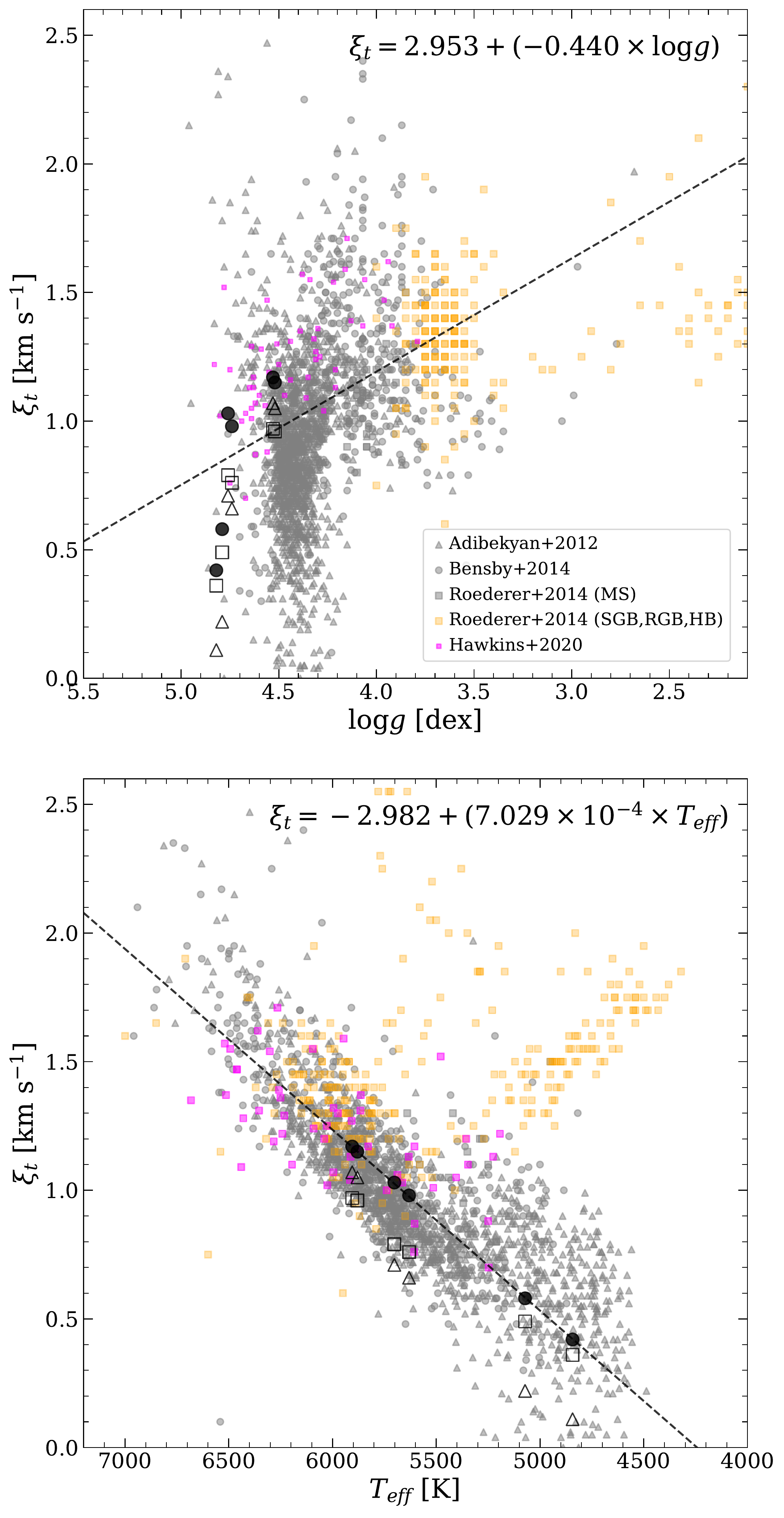}
     \caption{ $\xi_{t}$-$\log{g}$ (top panel) and $\xi_{t}$-$T_{\rm eff}$ (bottom panel) correlations for the literature data, which are 
     FGK dwarf stars of \citet[][grey triangles]{Adibekyan2012}, FG dwarf stars of \citet[][grey circles]{Bensby2014}, 
     main-sequence to horizontal-branch stars of \citet[][grey and orange squares]{Roederer2014}, 
     and main-sequence stars of \citet[][magenta squares]{Hawkins2020}.
     The dashed lines represent the least-squares fit of the data excluding non-main-sequence stars of \citet{Roederer2014} and stars 
     with $T_{\rm eff}$ $<$ 5000 K of \citet{Adibekyan2012} and \citet{Bensby2014}. 
     The $\xi_{t}$ has a better correlation with $T_{\rm eff}$ than $\log{g}$. 
    {Black circles are our target stars with $\xi_{t}$ values obtained from Equation~(\ref{eq:vt}). Open triangles and squares shows different $\xi_{t}$ values estimated using the equations of \citet{Boeche2016} and \citet{Mashonkina2017}, respectively.}
     }
     \label{fig:vt}
\end{figure}

In order to estimate the microturbulence velocity ($\xi_{t}$), we derived the $\xi_{t}-\log{g}$ and $\xi_{t}-T_{\rm eff}$ correlations for main-sequence stars based on a collection of literature data (see Figure~\ref{fig:vt}) as follows: 
\begin{eqnarray}
\xi_{t} = (-0.440 \pm 0.036) \times \log{g} + (2.953 \pm 0.157), \\
\xi_{t} = (7.029 \pm 0.155) \times 10^{-4} \times T_{\rm eff} - (2.982 \pm 0.089). \label{eq:vt}
\end{eqnarray}
The $\xi_{t}$ for our targets were then calculated from the $T_{\rm eff}$, which shows a better correlation with $\xi_{t}$ than $\log{g}$.
In addition, we compared the obtained $\xi_{t}$ values to those estimated using the relations of \citet{Boeche2016} and \citet{Mashonkina2017} in Table~\ref{tab:vt} and Figure~\ref{fig:vt}. 
Although our values are somewhat higher than the others, the difference in [Fe/H] by varying $\xi_{t}$ by these typical deviations is smaller than 0.05 dex for each star.

\begin{table}
\caption{Microturbulence velocity estimated using different equations}
\label{tab:vt} 
\centering                                    
\begin{tabular}{c c c c} 
\hline\hline                        
\multirow{2}{*}{ID}     & \multicolumn{3}{c}{$\xi_{t}$ [km~s$^{-1}$]} \\
                                & Eq. (3) & BG16 & M17 \\ 
\hline                                   
WB1a & 1.03 & 0.71 & 0.79 \\ 
WB1b & 0.98 & 0.66 & 0.76 \\
WB2a & 0.58 & 0.22 & 0.49 \\
WB2b & 0.42 & 0.11 & 0.36 \\
WB3a & 1.17 & 1.07 & 0.97 \\
WB3b & 1.15 & 1.05 & 0.96 \\
\hline                                             
\end{tabular}
\tablefoot{Each column indicates $\xi_{t}$ values estimated using Equation~(\ref{eq:vt}) in this study, Equations (A.1) and (A.2) in \citet{Boeche2016}, and Equation (2) in \citet{Mashonkina2017}.}  
\end{table}

We first estimated [Fe/H] from equivalent widths (EWs) of Fe lines with the  $T_{\rm eff}$,  $\log{g}$, and $\xi_{t}$ derived above. 
Then we iteratively performed the above procedures with the obtained [Fe/H] as the input value in the estimation of $T_{\rm eff}$ until the newly estimated [Fe/H] was the same as the input value.

\begin{figure}
\centering
   \includegraphics[width=0.43\textwidth]{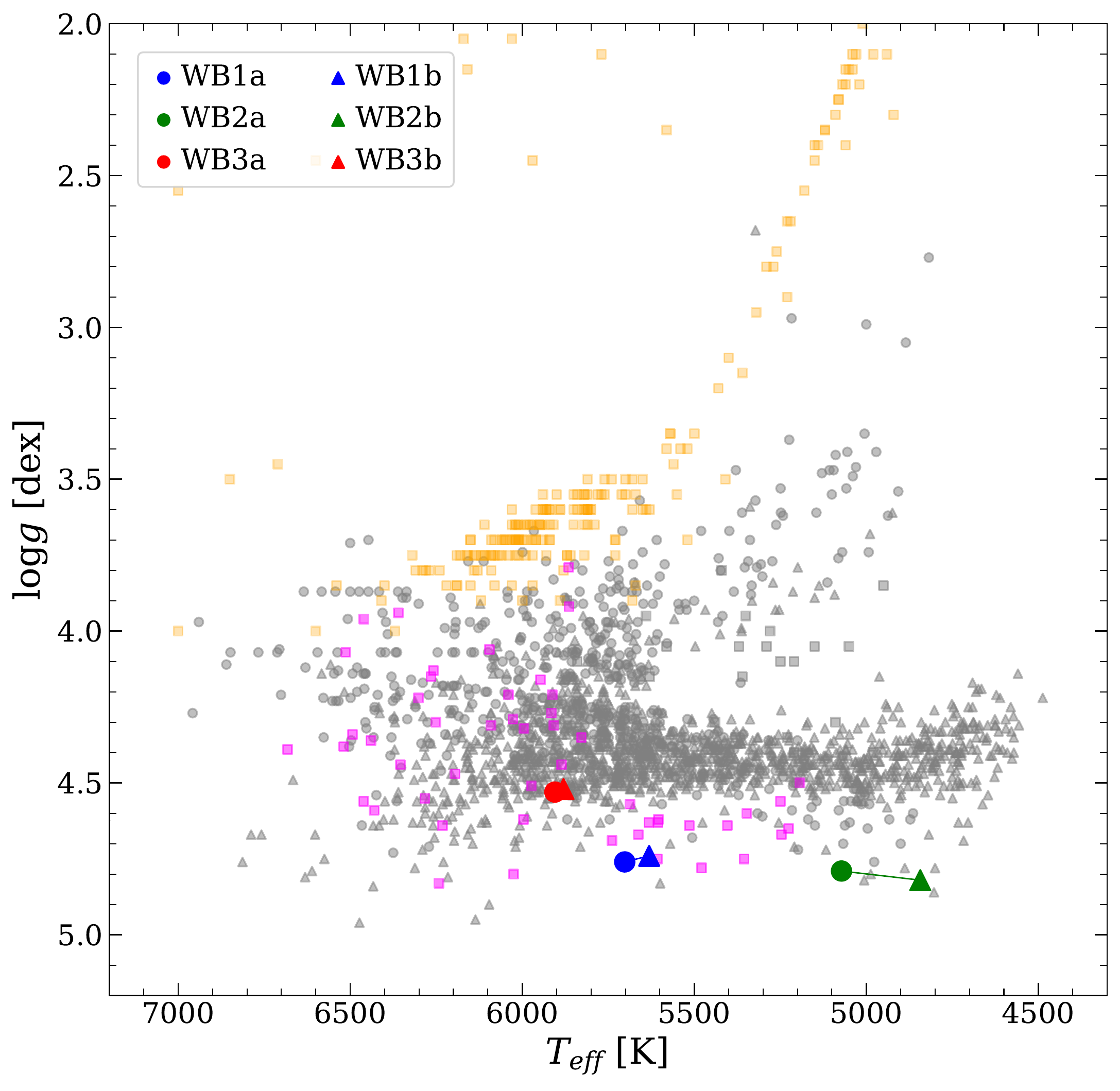}
     \caption{Our wide binary samples and their location on a Kiel diagramm, together with literature data as in Figure~\ref{fig:vt}.
     The WB1 (blue) and WB3 (red) pairs show very similar effective temperature and surface gravity between the component stars ($\Delta T_{\rm eff}$ $\le$ 75~K; $\Delta \log{g}$ $\le$ 0.02 dex),
     while WB2 (green) has a larger difference in the effective temperature ($\Delta T_{\rm eff}$ = 229~K). 
     }
     \label{fig:Teff_logg}
\end{figure}

The finally obtained atmosphere parameters yield a balance of the Fe-abundance with excitation potential, but an equilibrium with reduced EW could not be reached.
Although we attempted to derive spectroscopic $\log{g}$ for the sample stars from the ionisation equilibrium between Fe\,{\sc i}  and Fe\,{\sc ii}, we did not employ this value due to the high uncertainty on the EW measurements of Fe\,{\sc ii} lines.
For $\xi_{t}$, we could not derive realistic values to reduce the trend between abundance and reduced EW at given $T_{\rm eff}$ and $\log{g}$.
We also note that $T_{\rm eff}$ values derived from (V$-$K$_{\rm s}$) colour using the equation and coefficients of \citet{GonzalezHernandez2009} differ by up to 350 K from those using (BP$-$RP) colour.
These discrepancies lead to a difference of up to 0.2 dex  in [Fe/H] for each star. In this study, we employed the Gaia (BP$-$RP) colour to estimate $T_{\rm eff}$, based on their precise photometry.
Our sample stars are plotted in a Kiel diagram in Figure~\ref{fig:Teff_logg}, and their atmospheric parameters are listed in Table~\ref{tab:param}.

\begin{table}
\caption{Heliocentric velocities and atmospheric parameters}
\label{tab:param} 
\setlength{\tabcolsep}{5pt}
\centering                                    
\begin{tabular}{c c c c c c} 
\hline\hline                        
\multirow{2}{*}{ID}     & v$_{\rm HC}$          & $T_{\rm eff}$ & $\log{g}$     & $\xi_{t}$                       & [Fe/H]        \\
                                & [km~s$^{-1}$] & [K]                   & [dex]           & [km~s$^{-1}$] & [dex]         \\
\hline                                   
WB1a & $-$71.13 $\pm$ 0.3 & 5702 & 4.76 & 1.03 & -1.16 \\
WB1b & $-$70.97 $\pm$ 0.1 & 5631 & 4.74 & 0.98 & -1.16 \\
WB2a & 311.34 $\pm$ 1.2 & 5072 & 4.79 & 0.58 & -1.43 \\
WB2b & 311.88 $\pm$ 0.9 & 4843 & 4.82 & 0.42 & -1.41 \\
WB3a & $-$66.22 $\pm$ 0.9 & 5905 & 4.53 & 1.17 & -0.78 \\
WB3b & $-$65.40 $\pm$ 3.5 & 5880 & 4.52 & 1.15 & -0.80 \\
\hline                                             
\end{tabular}
\end{table}

\subsection{Chemical abundance analysis}
For spectroscopic abundance analysis, we applied model atmospheres interpolated from the ATLAS9 grid with $\alpha$-enhanced opacity distributions (AODFNEW) by \citet{Castelli2003}.
The chemical abundances were determined using the 2017 version of the local thermodynamic equilibrium (LTE) code MOOG \citep{Sneden1973}.
We used the {\em abfind} driver of MOOG to derive Fe, Na, Si, Ca, Sc, Ti, V, Cr, Mn, Co, Ni, Cu, Zn, Y, Zr, and Ba abundances from the EWs of absorption lines, while the Mg abundance was obtained from spectrum synthesis using the MOOG {\em synth} driver.
The EWs were measured by fitting single or multiple Gaussian profiles to isolated and blended lines with the semi-automatic code developed by \citet{Johnson2014}.
Table~\ref{tab:line} presents the line list used in this work, which is based on that of \citet{Koch2014}.
Hyperfine splitting was accounted for elements in which the splitting is significant (see the notes in Table~5). 
We note, however, that EWs of several lines within our spectral coverage for other elements, such as Al and Eu, could not be measured because the S/N was insufficient. 
The chemical abundances for each element, their line-to-line scatter ($\sigma$), and the number of lines used (N) are listed in Table~\ref{tab:abund} in terms of [X/H], adopting the solar abundance of \citet{Asplund2009}. 

\begin{table}
\caption{Line list}
\label{tab:line} 
\centering                                    
\begin{tabular}{c c c c} 
\hline\hline 
Wavelength      & Species       & EP            & $\log{gf}$ \\
$[\AA]$         &               & [eV]  &  \\
\hline 
5682.63 & Na\,{\sc i} & 2.102 &  $-$0.700 \\
5688.20 & Na\,{\sc i} & 2.105 &  $-$0.460 \\
5528.42 & Mg\,{\sc i} & 4.346 &  $-$0.481 \\
5711.09 & Mg\,{\sc i} & 4.330 &  $-$1.660 \\
5684.48 & Si\,{\sc i} & 4.954 &  $-$1.650 \\
5948.55 & Si\,{\sc i} & 5.083 &  $-$1.298 \\
6155.13 & Si\,{\sc i} & 5.610 &  $-$0.400 \\
\hline
\end{tabular}
\tablefoot{The full table is available at the CDS.}
\end{table}

\begin{table*}
\setlength{\tabcolsep}{4pt}
\caption{Chemical abundance results} 
\label{tab:abund} 
\centering                                    
\begin{tabular}{c r r r | r r r | r r r | r r r | r r r | r r r} 
\hline\hline 
\multirow{2}{*}{Species}        & \multicolumn{3}{c}{WB1a} & \multicolumn{3}{c}{WB1b} & \multicolumn{3}{c}{WB2a} & \multicolumn{3}{c}{WB2b} & \multicolumn{3}{c}{WB3a} & \multicolumn{3}{c}{WB3b} \\
        & [X/H] & $\sigma$ & $N$ & [X/H] & $\sigma$ & $N$ & [X/H] & $\sigma$ & $N$ & [X/H] & $\sigma$ & $N$ & [X/H] & $\sigma$ & $N$ & [X/H] & $\sigma$ & $N$ \\
\hline 
Fe\,{\sc i} & $-$1.16 & 0.17 & 85 & $-$1.16 & 0.17 & 83 & $-$1.43 & 0.15 & 88 & $-$1.41 & 0.17 & 90 & $-$0.78 & 0.19 & 77 & $-$0.80 & 0.16 & 89 \\
Na\,{\sc i} & $-$1.38 & 0.11 & 2 & $-$1.19 & 0.11 & 2 & $-$1.84 & 0.01 & 2 & $-$1.87 & 0.07 & 2 & $-$0.61 & \ldots & 1 & $-$0.73 & 0.13 & 2 \\
Mg\,{\sc i}$^{S}$ & $-$0.77 & 0.04 & 2 & $-$0.82 & 0.04 & 2 & $-$1.20 & 0.01 & 2 & $-$1.23 & 0.05 & 2 & $-$0.51 & 0.07 & 2 & $-$0.43 & 0.09 & 2 \\
Si\,{\sc i} & $-$1.08 & 0.38 & 2 & $-$0.82 & 0.05 & 2 & \ldots & \ldots & \ldots & \ldots & \ldots & \ldots & $-$0.70 & 0.65 & 2 & $-$0.66 & 0.29 & 3 \\
Ca\,{\sc i} & $-$0.81 & 0.13 & 15 & $-$0.78 & 0.14 & 16 & $-$1.17 & 0.14 & 13 & $-$1.23 & 0.16 & 14 & $-$0.52 & 0.18 & 15 & $-$0.52 & 0.15 & 16 \\
Sc\,{\sc ii}$^{H}$ & $-$1.11 & 0.16 & 2 & $-$1.17 & 0.11 & 2 & $-$1.51 & \ldots & 1 & $-$1.46 & 0.56 & 2 & $-$0.56 & 0.20 & 2 & $-$0.57 & 0.13 & 4 \\
Ti\,{\sc i} & $-$0.68 & 0.23 & 10 & $-$0.71 & 0.20 & 11 & $-$1.31 & 0.18 & 11 & $-$1.34 & 0.16 & 10 & $-$0.68 & 0.21 & 9 & $-$0.54 & 0.18 & 11 \\
V\,{\sc i} & $-$0.47 & \ldots & 1 & $-$0.55 & \ldots & 1 & $-$1.46 & \ldots & 1 & $-$1.38 & 0.02 & 2 & $-$0.61 & 0.09 & 2 & $-$0.60 & 0.24 & 3 \\
Cr\,{\sc i} & $-$1.02 & 0.16 & 6 & $-$1.03 & 0.05 & 6 & $-$1.22 & 0.17 & 10 & $-$1.43 & 0.11 & 6 & $-$0.91 & 0.13 & 9 & $-$0.82 & 0.05 & 6 \\
Mn\,{\sc i}$^{H}$ & $-$1.46 & 0.19 & 2 & $-$1.62 & 0.15 & 3 & $-$1.83 & 0.17 & 5 & $-$1.72 & 0.05 & 5 & $-$1.25 & \ldots & 1 & $-$1.18 & 0.21 & 5 \\
Co\,{\sc i}$^{H}$ & $-$0.84 & \ldots & 1 & $-$0.94 & 0.14 & 2 & $-$1.30 & \ldots & 1 & $-$1.26 & 0.28 & 2 & \ldots & \ldots & \ldots & $-$0.79 & \ldots & 1 \\
Ni\,{\sc i} & $-$1.03 & 0.24 & 11 & $-$1.17 & 0.24 & 9 & $-$1.50 & 0.24 & 11 & $-$1.50 & 0.25 & 10 & $-$0.85 & 0.17 & 7 & $-$0.82 & 0.16 & 8 \\
Cu\,{\sc i}$^{H}$ & \ldots & \ldots & \ldots & $-$1.40 & 0.09 & 2 & \ldots & \ldots & \ldots & $-$2.24 & \ldots & 1 & $-$1.04 & 0.09 & 2 & $-$0.92 & 0.08 & 2 \\
Zn\,{\sc i} & $-$0.97 & 0.15 & 2 & $-$0.74 & 0.23 & 2 & $-$1.21 & 0.16 & 2 & \ldots & \ldots & \ldots & $-$1.08 & 0.20 & 2 & $-$0.89 & 0.25 & 2 \\
Y\,{\sc ii} & $-$1.05 & 0.28 & 3 & $-$0.88 & 0.39 & 3 & $-$1.66 & 0.19 & 2 & $-$1.40 & 0.30 & 2 & $-$0.75 & 0.04 & 2 & $-$0.70 & 0.12 & 4 \\
Zr\,{\sc ii} & $-$0.47 & \ldots & 1 & $-$0.52 & \ldots & 1 & \ldots & \ldots & \ldots & $-$0.51 & \ldots & 1 & $-$0.55 & \ldots & 1 & \ldots & \ldots & \ldots \\
Ba\,{\sc ii}$^{H}$ & $-$0.76 & 0.11 & 3 & $-$0.83 & 0.14 & 3 & $-$1.39 & 0.09 & 3 & $-$1.34 & 0.03 & 3 & $-$0.53 & 0.14 & 2 & $-$0.58 & 0.08 & 3 \\
\hline                                             
\end{tabular}
\tablefoot{$\sigma$ indicates the standard deviation in the line-by-line abundances. An S denotes that this abundance was derived from spectrum synthesis, and H indicates that hyperfine structure was accounted for in the abundance measurement.}
\end{table*}

\subsection{Error analysis} \label{sec:error}

\begin{table*}
\caption{Systematic error due to the uncertainty in the atmosphere parameters}
\label{tab:error} 
\centering                                    
\begin{tabular}{c c c c c c c c} 
\hline\hline 
\multirow{2}{*}{Species}        & $T_{\rm eff}$ & $\log{g}$     & [Fe/H]        & $\xi_{t}$               & \multirow{2}{*}{ODF}  & \multirow{2}{*}{$\sigma_{total}$}     & \multirow{2}{*}{NLTE} \\
                                        & $\pm$65 K     & $\pm$0.1 dex   & $\pm$0.17 dex & $\pm$0.13 km~s$^{-1}$ & & & \\
\hline 
Fe\,{\sc i}     & $\pm$0.06     & $\mp$0.02     & $\pm$0.04     & $\mp$0.02     & $-$0.04 & 0.09  & $+$0.01       \\
Na\,{\sc i}     & $\pm$0.04     & $\mp$0.01     & $\pm$0.00     & $\mp$0.00     & $+$0.01 & 0.04  & \ldots                \\
Mg\,{\sc i}     & $\pm$0.04     & $\mp$0.02     & $\pm$0.03     & $\mp$0.01     & $-$0.03 & 0.06  & $-$0.01       \\
Si\,{\sc i}     & $\pm$0.02     & $\pm$0.01     & $\pm$0.01     & $\mp$0.00     & $-$0.01 & 0.03  & $-$0.00       \\
Ca\,{\sc i}     & $\pm$0.05     & $\mp$0.02     & $\pm$0.02     & $\mp$0.01     & $-$0.01 & 0.06  & $-$0.00 \\
Sc\,{\sc ii}    & $\pm$0.00     & $\pm$0.04     & $\pm$0.05     & $\mp$0.01     & $-$0.07 & 0.10  & \ldots                \\
Ti\,{\sc i}     & $\pm$0.08     & $\mp$0.01     & $\pm$0.01     & $\mp$0.02     & $+$0.01 & 0.08  & $+$0.18       \\
V\,{\sc i}      & $\pm$0.08     & $\pm$0.00     & $\pm$0.00     & $\mp$0.00     & $+$0.02 & 0.08  & \ldots                \\
Cr\,{\sc i}     & $\pm$0.08     & $\mp$0.02     & $\pm$0.02     & $\mp$0.02     & $-$0.01 & 0.09  & $+$0.13       \\
Mn\,{\sc i}     & $\pm$0.06     & $\pm$0.00     & $\pm$0.01     & $\mp$0.01     & $+$0.00 & 0.06  & $+$0.19       \\
Co\,{\sc i}     & $\pm$0.04     & $\pm$0.01     & $\pm$0.02     & $\mp$0.00     & $-$0.02 & 0.05  & $+$0.25       \\
Ni\,{\sc i}     & $\pm$0.04     & $\pm$0.00     & $\pm$0.03     & $\mp$0.01     & $-$0.03 & 0.06  & \ldots                \\
Cu\,{\sc i}     & $\pm$0.05     & $\pm$0.01     & $\pm$0.02     & $\mp$0.01     & $-$0.02 & 0.06  & \ldots                \\
Zn\,{\sc i}     &$\pm$0.00      & $\pm$0.02     & $\pm$0.03     & $\mp$0.01     & $-$0.04 & 0.05  & \ldots                \\
Y\,{\sc ii}     & $\pm$0.01     & $\pm$0.04     & $\pm$0.05     & $\mp$0.01     & $-$0.07 & 0.10  & \ldots                \\
Zr\,{\sc ii}    & $\pm$0.01     & $\pm$0.04     & $\pm$0.04     & $\mp$0.00     & $-$0.06 & 0.08  & \ldots                \\
Ba\,{\sc ii}    & $\pm$0.02     & $\pm$0.02     & $\pm$0.06     & $\mp$0.04     & $-$0.10 & 0.13  & \ldots                \\
\hline                                             
\end{tabular}
\end{table*}

To obtain an estimate for the statistical error on the abundance ratios, we list in Table~5 the line-to-line scatter and the number of lines we used for each element.
Next, we estimated the systematic errors of the chemical abundance measurements, which can be caused by uncertainties in atmospheric parameters, through comparison with abundances measured from altered atmosphere models. 
For this purpose, we interpolated eight atmosphere models with different atmosphere parameters varied by the typical uncertainty of our samples from the finally determined values ($\Delta T_{\rm eff}$ = $\pm$65~K, $\Delta \log{g}$ = $\pm$0.1~dex, $\Delta$[Fe/H] = $\pm$0.17~dex, and $\Delta \xi_{t}$ = $\pm$0.13~km~s$^{-1}$), and another model calculated using scaled-solar opacity distribution (ODFNEW).

The above uncertainties were computed by Monte Carlo sampling accounting for the errors on each input variable in the estimation.
We first took the uncertainty in [Fe/H] from the standard deviation of the line-by-line Fe abundance measurements (see Table~\ref{tab:abund}). 
The uncertainty in $T_{\rm eff}$ was estimated, taking the above uncertainty in [Fe/H], magnitude error of $Gaia$ DR2, and the dispersion in the empirical calibration of \citet{Mucciarelli2020} into account.
In the same manner, we estimated uncertainties in $\log{g}$ and $\xi_{t}$ based on the uncertainties of stellar mass ($\pm$0.2~$M_{\odot}$), distance, magnitude, $T_{\rm eff}$, and fitting error of Equation~(\ref{eq:vt}).
Abundances were then re-derived with these altered models for all sample stars, and the systematic errors for each element depending on each parameter are listed in Table~\ref{tab:error}, averaged over all stars. 
The total systematic uncertainty is calculated to be the squared sum of all contributions, which should be seen as an upper limit owing to the covariances between the stellar parameters, however \citep[e.g.][]{McWilliam1995}. 

In addition, in order to examine the non-LTE effects on the determination of the abundances, we estimated the corrections  for lines with available data based on \citet{Bergemann2012, Bergemann2013, Bergemann2017}\footnote{http://nlte.mpia.de}. 
The average effects of non-LTE for Fe, Mg, Si, Ca, Ti, Cr, Mn, and Co elements are also given in Table~\ref{tab:error}, although these corrections were not applied to the interpretation of our spectroscopic results for the comparison with literature data (see Section~\ref{sec:tagging}).
The non-LTE effects for Fe, Mg, Si, and Ca are negligible compared to the systematic errors, while those for Ti, Cr, Mn, and Co are somewhat large.


\section{Abundance results} \label{sec:result}
\subsection{Chemical homogeneity of wide binaries} \label{sec:abund}
\begin{figure}
\centering
   \includegraphics[width=0.45\textwidth]{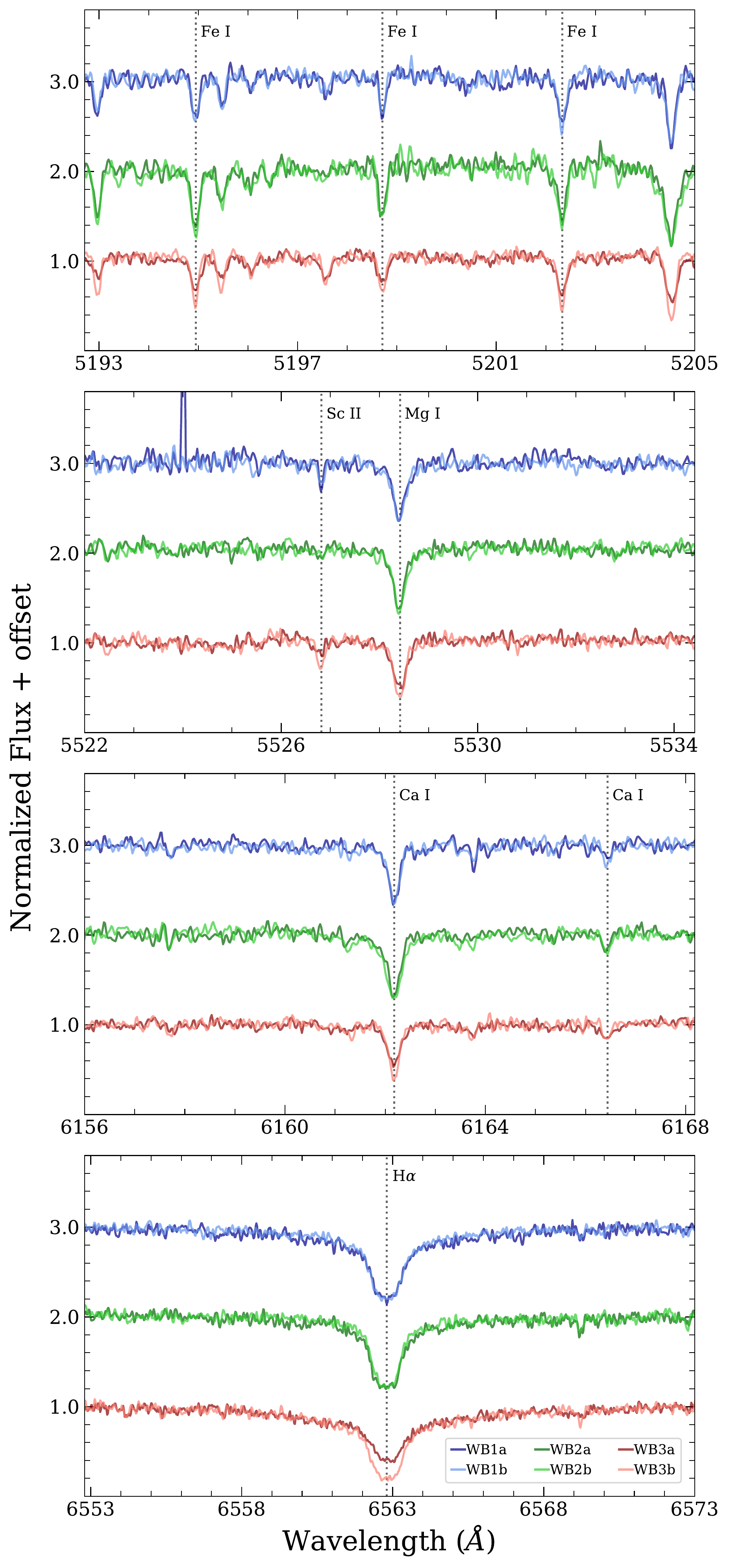}
     \caption{ 
     Continuum-normalised spectra of stars showing several Fe, Mg, Ca, and H$\alpha$ lines. 
     The spectra of the components in WB1 and WB2 are almost identical in these spectral ranges. 
     However, those for WB3 show differences in the strength of absorption lines, where WB3b has systematically stronger absorption lines.
     }
     \label{fig:spectra}
\end{figure}

For comparison, Figure~\ref{fig:spectra} shows observed spectra of components in each wide binary pair.
The stars of WB1 and WB2 have remarkably similar spectral features in the region of Fe, Mg, Ca, and H$\alpha$ absorption lines, suggesting a similar chemical composition of co-moving stars. 
In the case of WB3, however, the component WB3b has stronger absorption lines for these elements than WB3a.
It is intriguing that the WB3 pair shares similar abundances, in spite of the different shapes of parts of the spectra (see also Table~\ref{tab:abund}).  
In particular, the stronger H$\alpha$ absorption of WB3b is distinct and may indicate that WB3b is an unresolved binary itself, so that the continuum is falsified or the H$\alpha$ line would be affected by a hot dwarf companion.
If confirmed, the WB3 would be a good example for an emergence of wide binaries from higher-order multiplicity \citep[e.g.][]{Tokovinin2014,Halbwachs2017}.

\begin{figure*}
\centering
   \includegraphics[width=0.65\textwidth]{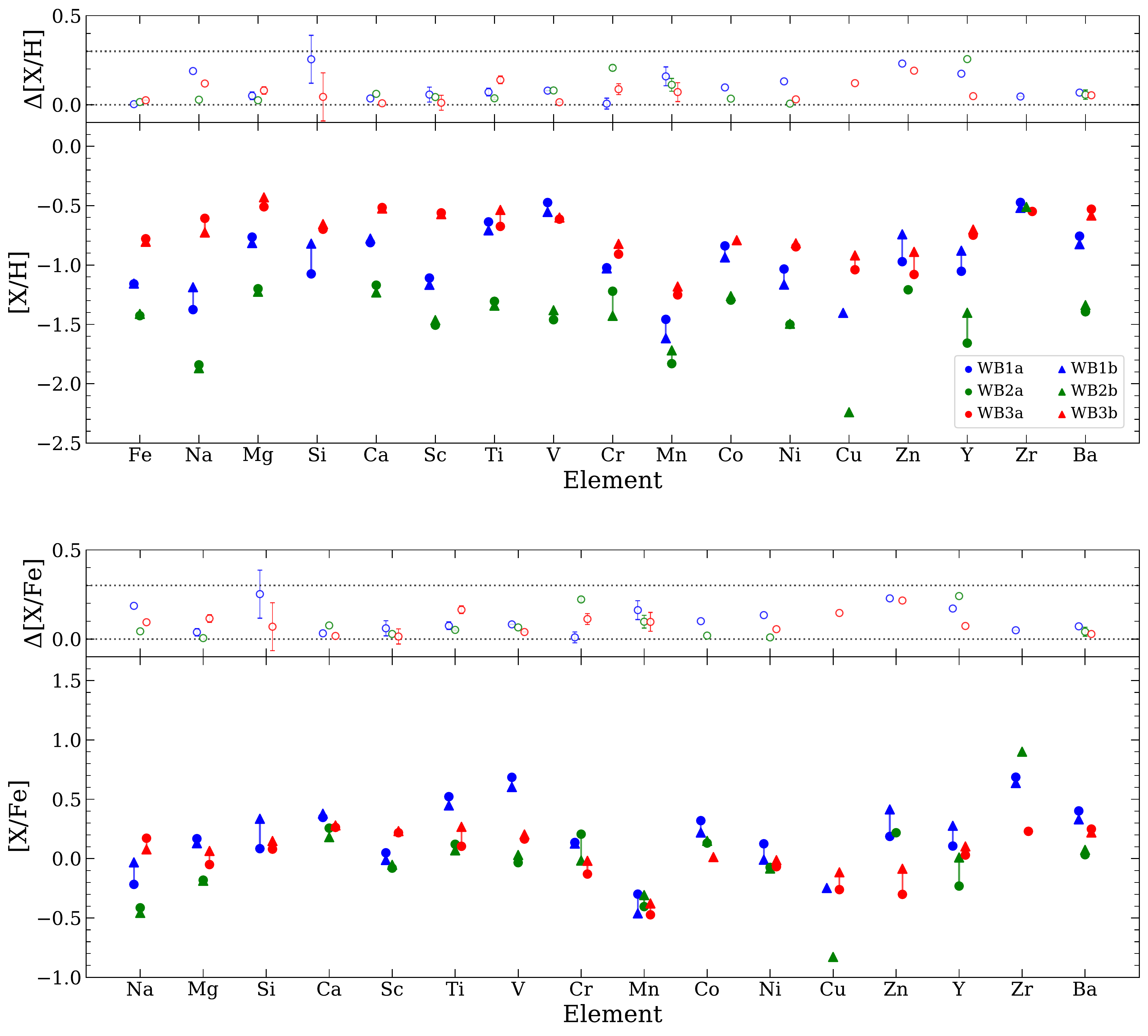}
     \caption{ 
     [X/H] (upper) and [X/Fe] (lower) abundance ratios of wide binary stars and their differences between component stars of each wide binary (blue for WB1, green for WB2, and red for WB3).
     Dotted lines in the $\Delta$[X/Y] plots denote the value of 0.0 and 0.3 dex, and the error bars indicate the 1$\sigma$ statistical error.  
     The differences in abundance between component stars are similarly small in the [X/H] and [X/Fe] abundance ratios. 
     }
     \label{fig:abund}
\end{figure*}

Our results show that the stars in each wide binary pair share very similar chemical abundances for all pairs we observed from our spectroscopic analysis (see Figure~\ref{fig:abund}).
In the case of [Fe/H], the differences between the component stars are only 0.003, 0.015, and 0.025 dex for WB1, WB2, and WB3, respectively, which is much smaller than the systematic uncertainty of 0.1 dex and the statistical error of $\sim$0.02 dex (taken as $\sigma$/$\sqrt{N}$; Table~\ref{tab:abund}).
The identical metallicity of the wide binary components, as well as 
close proper motion and line-of-sight velocity, therefore strongly support the hypothesis that our sample stars are indeed wide binary pairs and not randomly co-moving stars. 
The marginal variations of [Fe/H] in each wide binary are comparable to those found in the 
literature \citep[e.g.][]{Desidera2004,Desidera2006,Andrews2018,Hawkins2020}. 

Our wide binaries are placed on the metal-poor side compared to previously reported wide binaries. 
The mean values of [Fe/H] are $-$1.16, $-$1.42, and $-$0.79 dex for WB1, WB2, and WB3, respectively, while the majority of spectroscopically confirmed wide binaries in the literature are within the metallicity span of -1.0 $<$ [Fe/H] $<$ +0.5 \citep[e.g.][]{Hwang2021}.
This does not come as a surprise, however, because our wide binary samples can be dynamically associated with the Milky Way halo and have explicitly been selected as such (see the upper panel of Figure~\ref{fig:orbit}), whereas, for example, those of \citet{Andrews2018} and \citet{Hawkins2020} are associated with the Galactic disk.  
On the other hand, \citet{Hwang2021} have shown that the wide binary fraction is strongly dependent on metallicity, and it decreases with decreasing metallicity from [Fe/H] of 0.0 dex.
Therefore metal-poor wide binaries ([Fe/H] $<$ $-$1.0 dex) are expected to become extremely rare. 
We note that this metallicity dependence is mainly based on disk stars, and \citet{Hwang2021} claimed multiple formation channels of wide binaries depending on their metallicity.
In this regard, the WB2 pair with [Fe/H] of $-$1.42 dex can provide an important and unusual opportunity to study the nature of metal-poor wide binaries in the halo. 

Figure~\ref{fig:abund} presents the chemical abundance ratios for each element determined in this study in terms of [X/H] and [X/Fe], together with the differences between the component stars of wide binaries, $\Delta$[X/Y].
Typically, the differences in abundances between either component are less than 0.1 dex, with some exceptions, such as the Si abundance of WB1 at the highest 0.25 dex level, probably due to the large line-to-line abundance variation (see Table~\ref{tab:abund} and Figure~\ref{fig:abund}).
The Mg, Ca, and Ba abundances, for example, which have been derived  from prominent and well-determined absorption lines, are highly consistent in every wide binary pair ($\Delta\lesssim$ 0.05 dex). 
In addition, we could not find a peculiar chemical difference pattern between specific elements in our sample wide binaries.
It  therefore appears that the stars within any given wide binary share the same chemical composition, indicating that they have formed in the same environment. 
These chemical homogeneities of wide binary in various elements are comparable to what \citet{Andrews2018} and \citet{Hawkins2020} reported, while the discrepancies are somewhat larger than those seen in \citet{Hawkins2020}. 
As described above, this is probably because of the larger uncertainties of our abundance measurement due to the lower S/N ($\sim$25 per pixel) compared to that of \citet[][S/N $\sim$ 105 per pixel]{Hawkins2020}. 


\subsection{Chemodynamical tagging}\label{sec:tagging}  

Chemical tagging is one of the best ways to trace the origin of stars \citep{Freeman2002}. 
This technique is widely employed in Galactic archaeology when extensive spectroscopic observations are available. 
In particular, numerous stellar streams and substructures, such as $Gaia$-Enceladus and Sequoia, have been revealed and characterised in the Milky Way through recent chemodynamical studies \citep[e.g.][]{Helmi2018, Myeong2019, Ji2020, Horta2021, Koch-Hansen2021, Prudil2021}. 
In this vein, we compared the chemical and dynamical properties of our wide binaries with Galactic field stars to investigate their origin. 

\begin{figure}
\centering
   \includegraphics[width=0.46\textwidth]{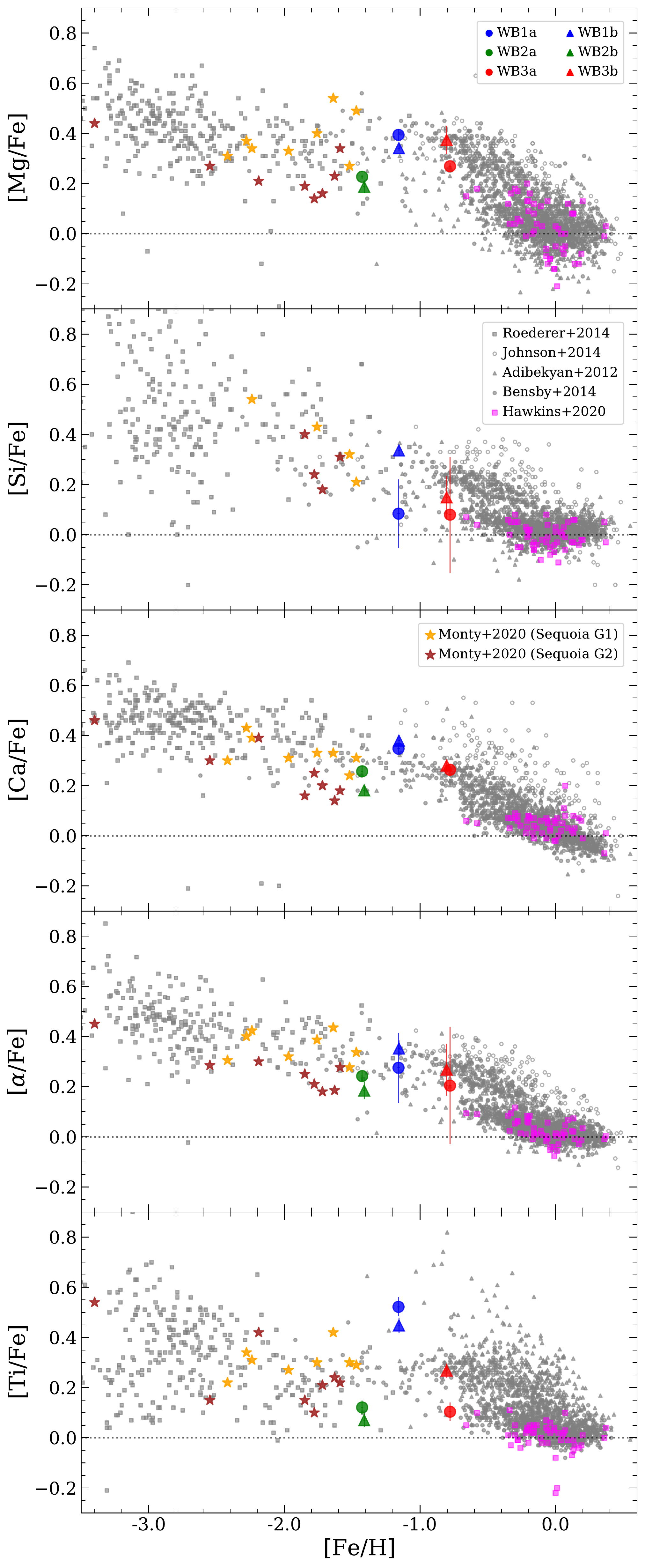}
     \caption{
     Chemical abundance ratios of wide binaries for individual and average $\alpha$-elements (Mg, Si, and Ca), and Ti, together with field stars in the Milky Way bulge \citep{Johnson2014}, disk \citep{Adibekyan2012, Bensby2014}, and halo \citep{Adibekyan2012, Roederer2014}. 
     We also plot wide binaries in the disk \citep[magenta squares:][]{Hawkins2020} and halo stars also associated with the Sequoia \citep[brown and yellow stars:][]{Stephens2002, Monty2020}.
     The horizontal dashed line indicates the solar level.
     It is interesting that the WB2 pair and the Sequoia G2 stars are commonly depleted in $\alpha$-elements compared to the general trend of field stars.
     }
     \label{fig:alpha}
\end{figure}

In Figure~\ref{fig:alpha} we plot the abundances of each $\alpha$-element (Mg, Si, and Ca), their average, and Ti for our sample stars in comparison with Galactic bulge, disks, and halo stars obtained from multiple literature sources (see caption for details). 
Interestingly, the WB2 pair shows lower $\alpha$-element abundances than the field stars within a similar metallicity region, in contrast to the other two pairs that closely follow their underlying the Milky Way halo environment. 
We note that the Si abundance of WB2ab stars could not be measured due to contamination by noise in the Si absorption lines. 
The large discrepancy in the Si abundance of the WB1 pair might also be affected by the large measurement error of WB1a (see Table~\ref{tab:abund} and Figure~\ref{fig:abund}).
The plot of the average $\alpha$-element abundance more clearly indicates the [$\alpha$/Fe] depletion of the WB2 pair. 
The [$\alpha$/Fe] abundances are 0.27/0.35 dex for WB1a/b, 0.24/0.18 dex for WB2a/b, and 0.20/0.27 dex for WB3a/b.
The abundance pattern of WB2 we derived is consistent with the earlier studies by \citet{King1997}, \citet{Chen2006}, \citet{Chen2014}, and \citet{Reggiani2018}.
These authors measured chemical abundances of this wide binary pair and reported [Fe/H] = $-$1.47 / $-$1.53 dex \citep{King1997} and $-$1.43 / $-$1.39 dex \citep{Reggiani2018} for WB2a ($\equiv$ HD134439) / WB2b ($\equiv$ HD134440), respectively, with an average [$\alpha$/Fe] of $\sim$ 0.03 dex.
This low $\alpha$-element abundance has already been interpreted as evidence of an accretion origin in both studies.
We note that the lower [$\alpha$/Fe] abundances of previous studies compared to ours are due to the significant depletion of [Mg/Fe], which is not shown in this study.

\begin{figure}
\centering
   \includegraphics[width=0.46\textwidth]{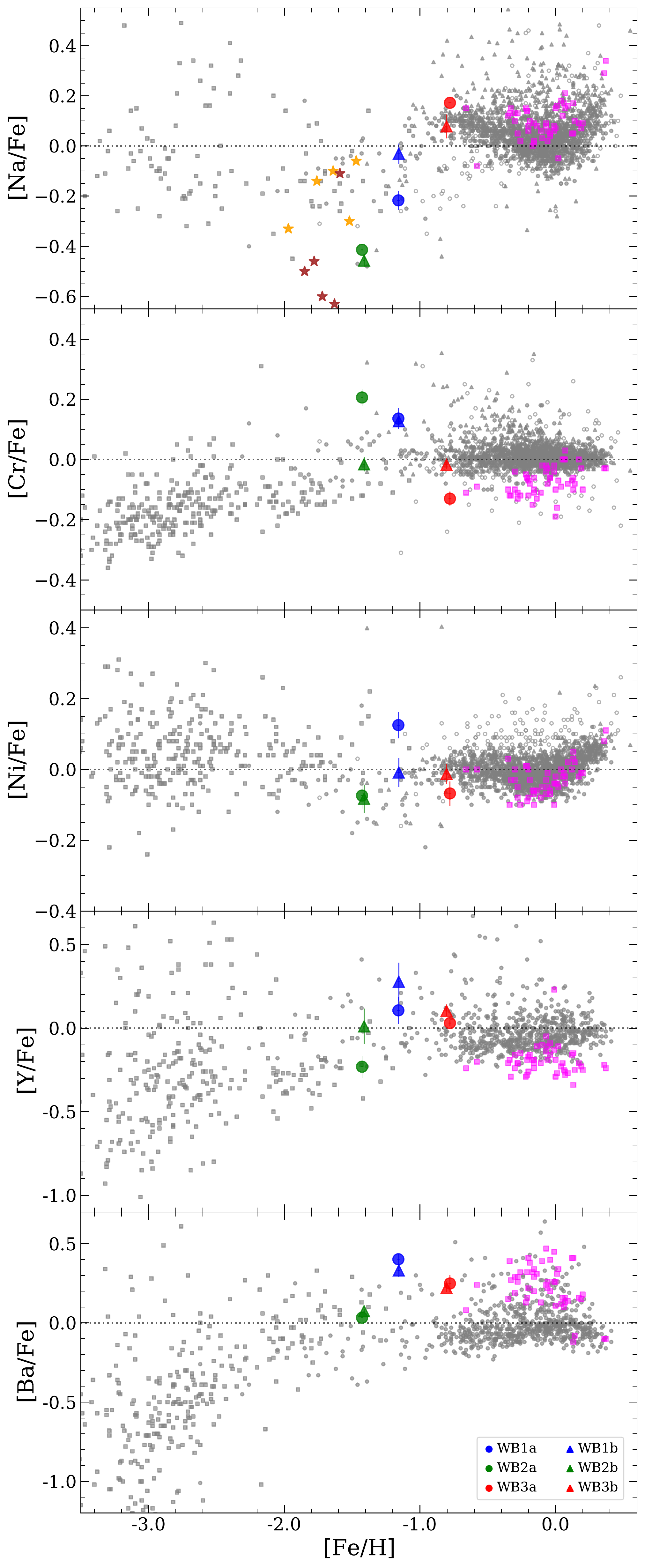}
     \caption{
     Same as Figure~\ref{fig:alpha}, but for Na, Cr, Ni, Y, and Ba. 
     }
     \label{fig:other}
\end{figure}

\begin{figure}
\centering
   \includegraphics[width=0.45\textwidth]{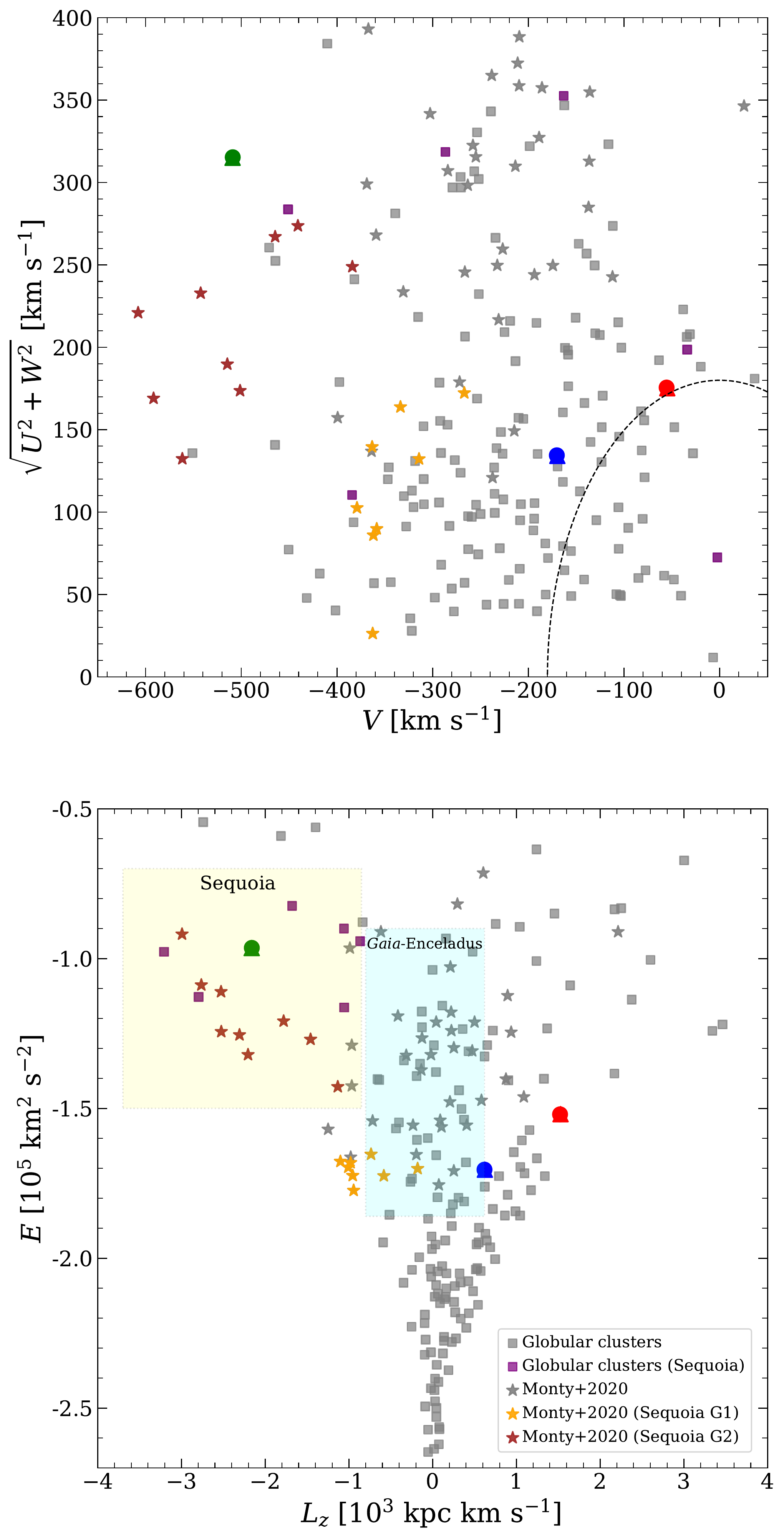}
     \caption{Toomre and $E$-$L_{Z}$ diagrams for our wide binary sample, globular clusters (grey squares), and stars from \citet[][stars]{Monty2020}. 
     Globular clusters associated with the Sequoia event are indicated as purple squares \citep{Massari2019}. 
     The curved dashed line in the upper panel denotes a total velocity of 
     180 km\,s$^{-1}$, which divides the halo and thick disk, 
     and yellow and cyan background boxes in the lower panel indicate the dynamical domains of the Sequoia and $Gaia$-Enceladus, respectively, as defined by \citet{Massari2019}. 
     Although the selection criterion for the Sequoia event slightly varies depending on the literature \citep[e.g.,][]{Koppelman2019, Myeong2019}, 
     the WB2 pair shows remarkably similar properties to the Sequoia stars and its globular clusters.
     }
     \label{fig:orbit}
\end{figure}

The low [$\alpha$/Fe] abundance ratio at a given [Fe/H] is an important chemical signature of stars that accreted from dwarf galaxies due to the low star-forming efficiencies of these low-mass systems \citep[see e.g.][]{Tinsley1979, Matteucci1990, Venn2004, Nissen2010, Hendricks2014, Hughes2020, Reichert2020}.
It is therefore likely that stars of WB2 were formed in a dwarf galaxy environment and then accreted into the Milky Way halo, as already conjectured by \citet{King1997}. 
Furthermore, the $\alpha$-element abundances of WB2 follow the trend of the Sequoia G2 stars defined by \citet{Monty2020}.
The Sequoia accretion event has been identified by \citet{Myeong2019} from the bulk of the high-energy retrograde stars and globular clusters in the halo \citep[see also][]{Koch2019b, Massari2019, Villanova2019}. 
The progenitor of Sequoia, with a predicted total mass between $\sim$10$^{8}~{M_{\odot}}$ \citep{Koppelman2019} and $\sim$10$^{10}~{M_{\odot}}$ \citep{Myeong2019}, was probably accreted onto the Milky Way 9--11 Gyr ago.
More recently, \citet{Monty2020} re-observed  stars of the \citet{Stephens2002} sample and  found that a number of stars are dynamically coincident with the $Gaia$-Sausage and Sequoia events, which were not known features in 2002. 
They also reported that stars originating from Sequoia could be divided into two groups, G1 and G2, with slightly different orbital energy and $\alpha$-element abundance patterns (see the yellow and brown stars in Figures~\ref{fig:alpha} -- \ref{fig:orbit}).

Figure~\ref{fig:other} shows chemical abundances ratios of Na, Cr, Ni, Y, and Ba  for our sample and field stars. 
It appears that the abundances of Fe-peak elements (Sc, V, Cr, Mn, Co, Cu, and Zn) and neutron-capture elements (Y, Zr, and Ba) of WB2 are on the global trend of the Milky Way stars, while slightly depleted in Ni abundance.  
This is difficult to confidently assert for these elements, however, owing to the large measurement and systematic errors. 
Nevertheless, we conclude that the WB2 pair and the Sequoia G2 stars also have comparable abundance ratios of [Na/Fe], [Y/Fe] and [Ba/Fe] \citep[see also Figure~13 of][]{Monty2020}. 

In order to further investigate the origin of WB2 and the binary sample, we estimated their orbital parameters, namely orbital energy ($E$) and angular momentum ($L$), of stars using the {\em galpy} package \citep{Bovy2015} with the Galactic potential of \citet{McMillan2017}. 
Figure~\ref{fig:orbit} presents the dynamical properties of our sample stars on a Toomre diagram (upper panel) and in the $E$-$L_{Z}$ plane (lower panel).  
WB2 is located in a region that is remarkably similar to that of the Sequoia G2 stars, not only in the Toomre diagram, but also in the $E$-$L_{Z}$ diagram. 
In addition, the orbital parameters of WB2 are also comparable to those of globular clusters that have been suggested  to be associated with the Sequoia event by \citet{Massari2019}. 
The dynamical properties of WB2 as well as its low $\alpha$-element abundances therefore strongly suggest that this wide binary  originated from an accreted dwarf galaxy, specifically, the progenitor of Sequoia. 

In contrast to WB2, we could not detect any obvious chemical nor dynamical peculiarities of the WB1 and WB3 binaries when compared to the Milky Way field stars.
Although the orbital parameters of WB1 mildly overlap with the $Gaia$-Enceladus event in the $E$-$L_{Z}$ plane, stars of this binary pair do not show a low [$\alpha$/Fe] abundance, which has been observed in the $Gaia$-Enceladus stars, however \citep[see e.g. Figure~4 of][]{Koppelman2019}.


\section{Discussion} \label{sec:discussion}
We confirm through spectroscopic observations that three wide pairs of common proper motion stars with high space velocities are indeed bound system, based on their common dynamical and chemical properties between the component stars. 
These stars are rare cases of wide binaries as metal-poor halo tracers \citep[see e.g.][]{Hwang2021}.
Our analysis demonstrates chemical homogeneity between their components, as is also seen in metal-rich wide binaries in the disks \citep[e.g.][]{Hawkins2020}.
This makes the chemical composition one of the global properties of a wide binary and suggests a common origin in a chemically similar environment. 

Moreover, our study suggests that one of our wide binary pairs is associated with the Sequoia event. 
An accretion origin of the WB2 pair has long been suspected: \citet{King1997} suggested chemical similarity of this wide binary with the globular clusters Rup~106 and Pal~12, which in turn are associated with the Magellanic Clouds. \citet{Reggiani2018} claimed a link with dwarf spheroidal galaxies such as Fornax.
In this study, we find that the WB2 is more likely related to the Sequoia event based on both chemical and dynamical properties.
This pair provides a significant opportunity to examine not only the formation mechanism of wide binaries, but also the assembly process of the Milky Way halo. 
At first glance, the existence of this wide binary supports the formation scenario suggested by \citet{Penarrubia2021}, which predicts that a large number of ultra-wide binaries with separations $\gtrsim$ 0.1 pc  arises from the disruption of low-mass systems, such as streams of globular clusters, over several Gyr. 
We note, however, that the physical separation of WB2 ($s$ $\sim$ 0.04 pc) is smaller than those of \citet[][$s$ $>$ 0.1pc]{Penarrubia2021}, and the total mass of the Sequoia is likely much more massive \citep[M$_{\rm Seq}$ $\sim$ $10^{10}$ M$_{\odot}$;][]{Myeong2019} than the progenitors adopted in the simulations of \citet{Penarrubia2021}. 
 
On the other hand, within the framework of \citet{Kouwenhoven2010} and \citet{Moeckel2011}, it is conceivable that both WB2 component stars were individually formed in a stellar cluster belonging to the Sequoia progenitor and were subsequently bound into a binary during the early dissolution phase of the cluster. 
If the WB2 components were already bound in the Sequoia progenitor either through the above process, the dynamical unfolding \citep{Reipurth2012}, or the formation from adjacent stellar cores \citep{Tokovinin2017}, it is significant to know whether the system can survive during the tidal disruption of the progenitor.
Whether this is a likely scenario depends on the properties of the progenitor dwarf galaxy and the peak separation of the wide binary.
The estimation in Appendix~\ref{apen:estimation} suggests that wide binaries with a combined mass of $1~M_{\odot}$ and a semi-major axis $<$ 1.4 pc will survive in the tidal field that disrupts dwarf galaxies with a half-light radius of $\sim$100 pc or larger.
Thus, it seems a plausible scenario that the WB2 pair formed in the progenitor galaxy and was later accreted onto the Galactic halo, although detailed dynamical modelling of the Sequoia progenitor and its disruption in the Milky Way tidal field is necessary to explore this possibility further.
In this regard, wide binaries may provide novel constraints on the hierarchical build up of the Milky Way halo.

A key remaining question is whether WB2 is a rare case or if a larger fraction of wide binaries can be associated with such Milky Way accretion events.
Interestingly, another peculiar wide binary (G 112-43/44) associated with the Helmi streams \citep{Helmi1999} has recently been reported by \citet{Nissen2021}.
This finding, along with ours, implies that more wide binaries related to accretion events are embedded in the Galactic halo.
In particular, because the WB2 is only $\sim$30 pc away from the Sun, we can reasonably expect that many more similar cases are waiting to be discovered in the entire Milky Way.
Although  WB2 is currently located in the Galactic disk region, its metallicity and dynamical properties rather associate it with the Galactic halo. 
This is also in line with the finding of an increasing separation with the binary height above the plane \citep{Sesar2008}.
Further searches for such wide binaries through chemodynamical studies will provide a novel opportunity to understand the formation of wide binaries and the assembly process of the Milky Way.

\vspace{5mm}
\begin{acknowledgements}
We thank the referee for a number of helpful suggestions.
DL and AJKH gratefully acknowledges funding by the Deutsche Forschungsgemeinschaft (DFG, 
German Research Foundation) -- Project-ID 138713538 -- SFB 881 (``The Milky Way System''), subprojects A03, A05, A11. 
DL thanks Sree Oh for the consistent support.
\end{acknowledgements}


\bibliographystyle{aa} 
\bibliography{export-bibtex} 


\appendix
\section{Survival or disruption of wide binary during the accretion event}\label{apen:estimation}
In the following, we estimate whether a wide binary in a progenitor dwarf galaxy could survive during the accretion of the system into the Milky Way.
The tidal radius of a dwarf satellite galaxy can be estimated as follows \citep{King1962}: 
\begin{eqnarray}
r_{\rm t} = \left(\frac{M_{\rm sat}}{M_{\rm MW}}\right)^{\frac{1}{3}} \times D, 
\end{eqnarray}
where $r_{\rm t}$, $M_{\rm sat}$, and $D$ are the tidal radius, mass, and Galactocentric distance of the dwarf satellite galaxy, and $M_{\rm MW}$ is the mass of the Milky Way. 
The satellite will lose its stars when $r_{t} < 2r_{\rm half,~sat}$, where $r_{\rm half,~sat}$ is the half-light radius of the satellite. 
In the same manner, the binary in the satellite will be disrupted if the tidal radius is
\begin{eqnarray}
r_{\rm t, WB} = \left(\frac{M_{\rm WB}}{M_{\rm MW}}\right)^{\frac{1}{3}} \times D < 2a,
\end{eqnarray}
where $M_{\rm WB}$ and $a$ are the mass and semi-major axis of binary.
Because the binary would experience the same tidal field as the affiliated dwarf galaxy, these two equations are combined based on the common terms. 
We further assume that the tidal interaction of the binary with the potential of the  satellite galaxy is negligible relative to that of the Milky Way.
Thus, 
we obtain the following condition for the disruption of a binary to occur:
\begin{eqnarray}
a > \left(\frac{M_{WB}}{M_{\rm sat}}\right)^{\frac{1}{3}} \times r_{\rm half,~sat}. \label{eq:con1}
\end{eqnarray}
In addition, \citet{Walker2009} found a tight correlation between mass and half-light radius of the dwarf spheroidal galaxy as follows: 
\begin{eqnarray}
M_{\rm sat} \sim M_{\rm sat}(r_{\rm half}) = 580~M_{\odot} \times \left(r_{\rm half,~sat}/{\rm pc}\right)^{1.4}. \label{eq:walker}
\end{eqnarray}
By employing Equation~(\ref{eq:walker}), Equation~(\ref{eq:con1}) can be converted into 
\begin{eqnarray}
a > 0.12 \times \left(\frac{M_{\rm WB}}{M_{\odot}}\right)^{\frac{1}{3}} \times (r_{\rm half,~sat}/{\rm pc})^{0.53}.
\end{eqnarray}
This equation implies that if we assume the dwarf satellite galaxy of $r_{\rm half} > 100$~pc, wide binaries of 1~$M_{\odot}$ with $a > 1.4$~pc will be disrupted upon accretion of the host system onto the Milky Way.
In other words, wide binaries with $a < 1.4$~pc could survive in the tidal field that disrupts the progenitor dwarf satellite galaxy. 
The condition of the semi-major axis for survival or disruption of wide binary  increases with increasing mass of the binary and the  half-light radius of the progenitor. 


\section{Further observations of binaries and non-binaries} \label{apen:other}

\begin{table*}
\small
\setlength{\tabcolsep}{4pt}
\caption{Details of the binary candidates and non-binaries that we did not use.}
\label{tab:append} 
\centering                                    
\begin{tabular}{ccccccccc} 
\hline\hline 
ID & $\alpha$ (J2000) & $\delta$ (J2000) & v$_{\rm HC}$  & $\mu_{\alpha}$        & $\mu_{\delta}$      & Parallax    & Ang. Sep &  Phy. Sep \\
 &  [hh:mm:ss] & [dd:mm:ss]  & [km~s$^{-1}$] &  [mas~yr$^{-1}$] &  [mas~yr$^{-1}$] & [mas] & [$\arcsec$] & [pc] \\
\hline
\multicolumn{9}{c}{MIKE}\\
\hline
NLTT~8753          & 02 42 05.05 & $-$24 45 23.1 &   70.6$\pm$1.5     & $-$69.866 & $-$421.636 & 5.4985 & &  \\
NLTT~8759          & 02 42 14.89 & $-$24 44 24.8 &   70.5$\pm$2.5     & $-$67.721 & $-$421.010 & 5.6226 & \raisebox{1.5ex}[-1.5ex]{146.2} & \raisebox{1.5ex}[-1.5ex]{0.13} \\                      
NLTT~16394         & 06 19 40.25 & $-$30 42 05.8 &  269.5$\pm$0.3     &    \phantom{$-$}305.161 & $-$163.910 & 5.5400 & & \\                                                          
NLTT~16407         & 06 20 27.24 & $-$30 36 18.1 &  269.8$\pm$1.0     &    \phantom{$-$}324.9 & $-$162.6 &  -- & \raisebox{1.5ex}[-1.5ex]{699.0} &  \raisebox{1.5ex}[-1.5ex]{0.61} \\ 
\hline
\multicolumn{9}{c}{MagE}\\
\hline
NLTT~11288         & 03 34 21.06 & $-$24 04 56.2 &    34.7$\pm$3.0    & $-$50.325 & $-$500.823 & 7.2622 & & \\
NLTT~11300         & 03 34 38.35 & $-$24 04 31.0 &    35.5$\pm$3.8    & $-$49.900 & $-$500.720 & 7.2441 & \raisebox{1.5ex}[-1.5ex]{223.5} & \raisebox{1.5ex}[-1.5ex]{0.15} \\
NLTT~30792         & 12 27 46.69 & $+$13 36 32.7 & $-$93.9$\pm$3.9    &    \phantom{$-$}73.794 & $-$265.891 & 2.4533 & & \\
NLTT~30795         & 12 27 43.85 & $+$13 34 12.0 & $-$98.0$\pm$2.2    &    \phantom{$-$}73.115 & $-$266.345 & 2.6027 & \raisebox{1.5ex}[-1.5ex]{146.6} &  \raisebox{1.5ex}[-1.5ex]{0.28}\\
NLTT~33984         & 13 23 55.74 & $+$20 26 30.4 &  $-$0.5$\pm$3.4    & $-$105.409 & $-$227.060 & 6.5849 & & \\                                                                                
NLTT~34019         & 13 24 30.49 & $+$20 27 18.8 &    68.6$\pm$3.7    & $-$94.573 & $-$203.187 & 2.8522 & \raisebox{1.5ex}[-1.5ex]{490.8} & \raisebox{1.5ex}[-1.5ex]{0.50} \\
NLTT~37787         & 14 34 44.72 & $+$25 06 56.3 &   373.9$\pm$4.1    & $-$45.465 & $-$342.347 & 5.1140 & & \\
NLTT~37790         & 14 34 51.06 & $+$25 09 57.8 &   371.8$\pm$2.9    & $-$45.680 & $-$342.395 & 5.1547 & \raisebox{1.5ex}[-1.5ex]{200.8} & \raisebox{1.5ex}[-1.5ex]{0.19} \\
\hline  
\multicolumn{9}{c}{FIES}\\
\hline
PM\_I14319+3113    & 14 31 58.00 & $+$31 13 46.2 &  $-$73.2$\pm$0.5     &    $-$100.407 & $-$59.456 & 3.2852 &  & \\ 
PM\_I14320+3119    & 14 32 03.16 & $+$31 19 49.0 &  $-$151.4$\pm$1.8     &    $-$97.960 & $-$57.994 & 2.2453 & \raisebox{1.5ex}[-1.5ex]{368.7} & \raisebox{1.5ex}[-1.5ex]{0.65}\\ 
PM\_I20449-0140    & 20 44 59.53 & $-$01 41 00.2 &   $-$8.3$\pm$0.5  & $-$34.577 &  $-$179.282 & 2.5965 & &  \\
PM\_I20450-0140    & 20 45 01.20 & $-$01 40 58.5 &   $-$30.2$\pm$1.0  & $-$34.777 &  $-$179.618 & 2.5749 & \raisebox{1.5ex}[-1.5ex]{25.1} &  \raisebox{1.5ex}[-1.5ex]{0.05}\\
\hline                                             
\end{tabular}
\tablefoot{The astrometric parameters are updated by $Gaia$ eDR3 \citep{GaiaCollaboration2021}. However, the proper motion of NLTT~16407 is taken from \citet{Chaname2004} because no $Gaia$ measurements are available.
The NLTT~33984/34019, PM\_I14319+3113/14320+3119, and PM\_I20449-0140/20450-0140 pairs are refuted as wide binaries based on their significantly different radial velocities.}  
\end{table*}


An earlier set of observations was acquired at lower spectral resolution and the data were of lower quality, allowing for a mere kinematic study.
We briefly describe these data and list the radial velocities so that these candidates can be in- and excluded in future statistical analyses, for instance in the assessment of the clumpy dark matter distribution in the halo \citep[e.g.][]{Chaname2004,Quinn2009}.

\subsection{MIKE spectra}
We observed two further binary pairs with the Magellan Inamori Kyocera Echelle (MIKE) spectrograph at the 6.5 m Magellan2/Clay Telescope in November 2008 and July 2009 using a slit width of 0.7$\arcsec$  and a binning of 2$\times$2 CCD pixels in the spatial and spectral dimensions.
For details of the data and their reduction, we refer to \citet{Quinn2009}.
While taken at high resolution (R$\sim$40000), the spectra are hampered by unfortunately low S/N ($<$ 20 per pixel at H$\alpha$) that only permitted radial velocity measurements, but not the derivation of chemical abundance ratios.

\subsection{MagE spectra}
During three nights in July 2009, we acquired spectra of four further wide binary candidate pairs with the Magellan Echellette (MagE) spectrograph \citep{Marshall2008}. 
Our target selection for this campaign was based on the same criteria as outlined in Section~\ref{sec:target}. 
MagE  yields a much lower spectral resolution than MIKE, and we integrated for considerably shorter times and will primarily focus on velocity measurements for assessing the binarity of the stars from these data. 
In practice, MagE was used with its standard setting, that is, with a 1$\arcsec$ slit and a binning of 1$\times$1 in spectral and spatial directions. This yielded a resolving power of R$\sim$4100. 

The data were reduced with the IDL-based pipeline kindly provided by G. Becker. 
This pipeline is largely built on the same principles and techniques as are used within the MIKE pipeline \citep{Kelson2003}. 
As before, the basic reduction steps comprise flatfield corrections using both Xe-flash lamp exposures and dome flats, wavelength calibration through Th-Ar lamps that were taken between each pair of science exposures, and optimal extraction and sky subtraction. 

\subsection{FIES spectra}
Two further pairs have been targeted with the FIES instrument within the same campaign that provided the sample we analysed here;
accordingly, the observation and reduction procedures are identical to those outlined in Section~\ref{sec:target_obs}. 
As they show very different radial velocities, these stars have not been incorporated in our above chemodynamic analysis.

\subsection{Radial velocities and non-binarity}
Using IRAF's {\em fxcor} task, we cross-correlated the spectra against a synthetic template with stellar parameters similar to the target stars. 
The results are listed in Table~\ref{tab:append}.

Out of the 11 candidate pairs that we observed as part of our campaign, 3 (i.e. 27\%) could be refuted as physically bound systems based on their significantly ($\gg$10$\sigma$) different radial velocities.
\citet{Quinn2009} have impressively demonstrated the sensitivity of the predicted mass range of the clumpy dark matter in the halo in the form of MACHOs to the widest-separation binaries.  
In particular, \citet{Quinn2010} argued that the in- or exclusion of even a single wide binary pair at very large angular separation has dramatic consequences for the conclusions that can be drawn; the removal of a pair at the 1~pc level erases the constraints on the MACHO models at the 95\% level.
However, given the availability of systematic, large samples \citep[e.g.][]{Monroy-Rodriguez2014,Coronado2018,Hartman2020} and the small size of our preselected sample, we
defer a reanalysis of the statistics following \citet{Quinn2009} to a future work.

\end{document}